\documentclass[aps,pra,twocolumn,superscriptaddress,longbibliography,footinbib]{revtex4-1}  %
\usepackage{graphicx} 
\usepackage{float} 
\usepackage{dcolumn} 
\usepackage{bm,color}
\usepackage{amsmath,amssymb,dsfont,amstext,amsfonts}
\usepackage{extarrows}
\usepackage{xcolor}
\usepackage{wasysym}
\usepackage{mathtools}
\usepackage{bbold}
\usepackage[export]{adjustbox}
\usepackage{mathdots}
\usepackage{siunitx}
\usepackage[colorlinks=true,linkcolor=blue,urlcolor=blue,citecolor=blue]{hyperref}
\usepackage[normalem]{ulem} 
\usepackage{color}
\usepackage{multibib}
\newcites{latex}{\LaTeX-Literature}

\newcommand{\parag}[1]{\vspace{0.3em}{\it \textcolor{blue}{#1.}}---}

\begin{document}
\title{Manipulation and braiding of Weyl nodes using symmetry-constrained phase transitions}
\author{Siyu Chen}
\email{sc2090@cam.ac.uk}
\affiliation{TCM Group, Cavendish Laboratory, University of Cambridge,
J. J. Thomson Avenue, Cambridge CB3 0HE, United Kingdom}
\author{Adrien Bouhon}
\email{adrien.bouhon@su.se}
\affiliation{Nordic Institute for Theoretical Physics (NORDITA),
Stockholm, Sweden}
\author{Robert-Jan Slager}
\email{rjs269@cam.ac.uk}
\affiliation{TCM Group, Cavendish Laboratory, University of Cambridge,
J. J. Thomson Avenue, Cambridge CB3 0HE, United Kingdom}
\author{Bartomeu Monserrat}
\email{bm418@cam.ac.uk}
\affiliation{TCM Group, Cavendish Laboratory, University of Cambridge,
J. J. Thomson Avenue, Cambridge CB3 0HE, United Kingdom}
\affiliation{Department of Materials Science and Metallurgy, University of Cambridge, 27 Charles Babbage Road, Cambridge CB3 0FS, United Kingdom}

\begin{abstract}
Weyl semimetals are arguably the most paradigmatic form of a gapless topological phase. While the stability of Weyl nodes, as quantified by their topological charge, has been extensively investigated, recent interest has shifted to the manipulation of the location of these Weyl nodes for non-Abelian braiding. To accomplish this braiding it is necessary to drive significant Weyl node motion using realistic experimental parameter changes. We show that a family of phase transitions characterized by certain symmetry constraints impose that the Weyl nodes have to reorganise by a large amount, shifting from one high-symmetry plane to another. Additionally, for a subset of pairs of nodes with non-trivial Euler class topology, this reorganization can only occur through a braiding process with adjacent nodes. As a result, the Weyl nodes are forced to move a large distance across the Brillouin zone and to braid, all driven by small temperature changes, a process we illustrate with Cd$_2$Re$_2$O$_7$. Our work opens up routes to readily manipulate Weyl nodes using only slight external parameter changes, paving the way for the practical realization of reciprocal space braiding.
\end{abstract}
\maketitle

\parag{Introduction} The discovery of topological insulators (TIs)~\cite{Rmp1,Rmp2} has reinvigorated interest in band theory over the past decade. The initial ideas behind topological insulators have been extended to include different crystalline symmetries, resulting in a plethora of topological characterisations~\cite{Clas1, Clas2, Clas3, Wi2, Clas4, Clas5, Codefects2, HolAlex_Bloch_Oscillations, probes_2D,ShiozakiSatoGomiK, Chenprb2012, mSI, mtqc, magenticpaper, wiedersemi, subdimhigh,Ft1, Axion3, bouhon2019wilson}, and also to topological semimetals and superconductors~\cite{RevModPhys.90.015001, Armitage2018, PhysRevLett.115.187001, PhysRevB.79.094504}. Within these, Weyl semimetals represent an ultimate consequence of translational symmetry, allowing for locally stable band crossings that carry a topological charge. More recently, there has also been interest in braiding Weyl nodes residing in different band gaps in systems exhibiting reality conditions, as this process can result in configurations in which nodes in the same gap have similar charges. The resulting obstruction for the nodes to annihilate is characterized by a new type of multi-gap invariant known as Euler class~\cite{bouhon2019wilson,BJY_nielsen, bouhon2019nonabelian, Wu1273, bouhonGeometric2020, peng2021nonabelian, jiang2021observation, Eulerdrive}.

The practical use of Weyl nodes requires their manipulation with external parameters, and there is a growing body of literature in this direction. Examples include the use of stress and strain~\cite{Sie2019, Modegrain}, coupling to electromagnetism and light~\cite{Zhang2017,de_Juan_2017}, and disorder~\cite{pixley2016rare, BbcWeyl, pixley2021rare, phasediagweyl}. However, in these proposals the Weyl nodes only move by a small fraction of the Brillouin zone dimension with experimentally realistic parameter changes, making the manipulation of Weyl nodes impractical. Here, we explore a route to manipulating Weyl nodes that combines insights from materials science via temperature-driven structural changes with more fundamental theoretical insights that find their origin in symmetry. Namely, we show that in a particular class of structural phase transitions, Weyl nodes need to travel a long distance across the Brillouin zone to follow the relevant symmetry changes, and these large displacements are driven by small temperature changes. We first exemplify this concept using Cd$_2$Re$_2$O$_7$ as a material example, and we then provide a more fundamental perspective setting the stage for a general mechanism. This allows us to predict this behavior for a whole class of compounds and symmetries.

\parag{Structural phase transitions}
Cd$_2$Re$_2$O$_7$ is an intriguing material as it is the only known superconductor in the pyrochlore family~\cite{Jin2001, Sakai2001}. Interest in its superconductivity has led to extensive exploration of its phase diagram under different conditions~\cite{Yamaura2017}. As summarized in Fig.~\ref{fig:distortion}(a), Cd$_2$Re$_2$O$_7$ crystallizes in a structure of cubic symmetry (phase I, Fd$\bar{3}$m)~\cite{Donohue1965} at ambient temperature and pressure. The structure can be described as two interwoven pyrochlore lattices assembled from Re$_4$ and Cd$_4$ tetrahedra, respectively, where the centers of O$_6$ octahedra coincide with the centers of Re$_4$ tetrahedra. With gradual cooling, two low-temperature phases with higher conductance appear successively at T$_{s1}$ = $\sim$\SI{200}{\kelvin}~\cite{Hanawa2001,Castellan2002} and T$_{s2}$ = $\sim$\SI{120}{\kelvin}~\cite{Yamaura2002,Hiroi2002}, both exhibiting tetragonal symmetry but lacking a center of inversion (phase II, I$\bar{4}$m2 and phase III, I$4_1$22). Very recently, a magnetic torque experiment has revealed that the transition between phases II and III is mediated by an additional phase of point group D$_{4h}$ or D$_{2}$~\cite{Uji2021}. Given that this additional phase was not detected in previous experiments with the exact same setting but a relatively large temperature step of \SI{2.5}{\kelvin}~\cite{Matsubayashi2020}, the new phase should only exist in the small temperature range between T$_{s2}-\Delta$T and T$_{s2}$, where $\Delta$T is less than \SI{2.5}{\kelvin}.

\begin{figure}
     \centering
     \includegraphics[width=0.5\textwidth]{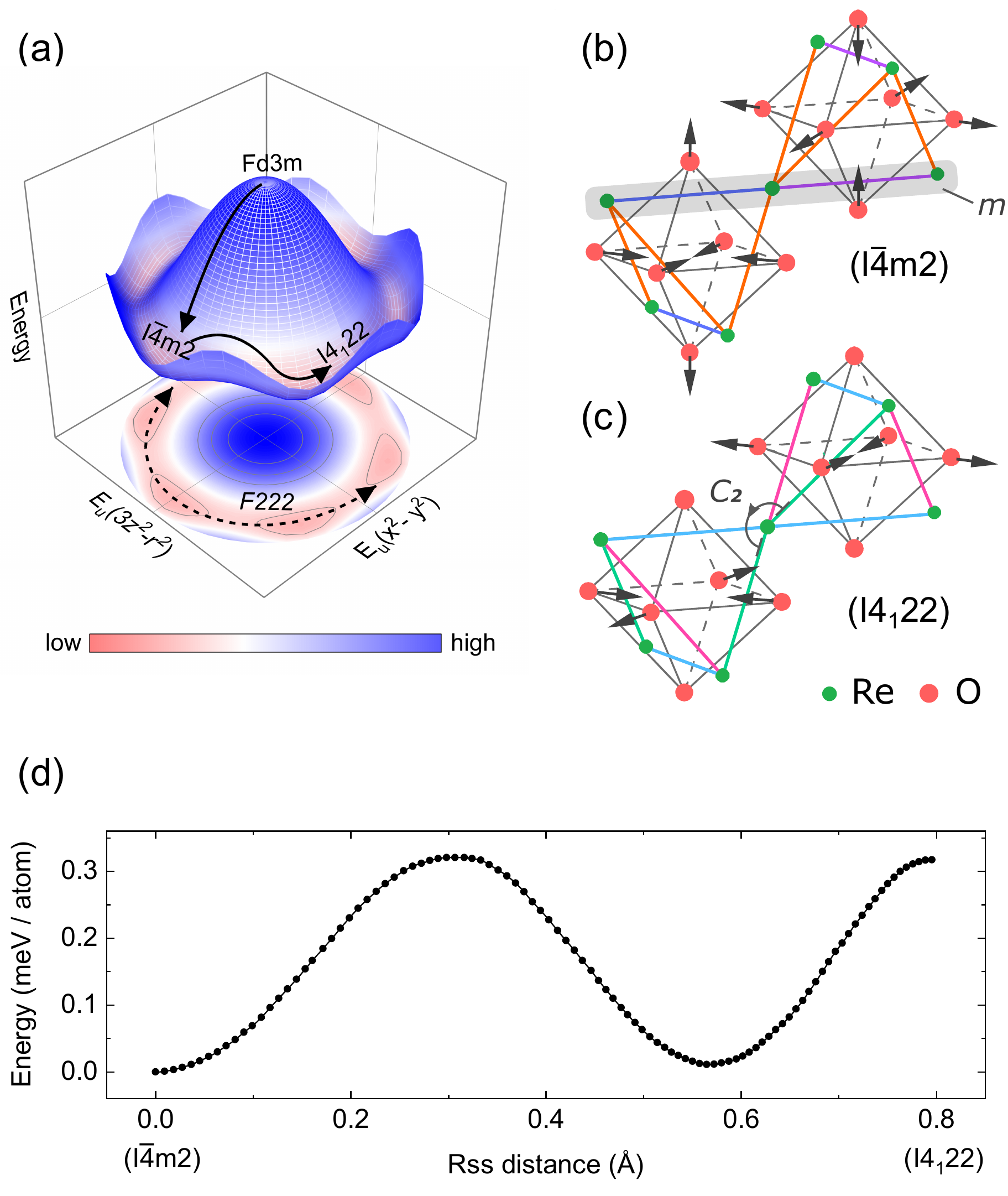}
     \caption{(a) Energy landscape of Cd$_2$Re$_2$O$_7$. The space groups of the three phases observed in experiments are shown on the energy surface. The solid arrows indicate the path of the phase transition during cooling and the dash arrow indicates the Goldstone mode fluctuating between phases II and III. (b-c) Distortions of neighbouring Re$_4$ tetrahedra and O$_6$ octahedra in (b) phase II and (c) phase III. The Re and O ions are represented by the green and red vertices respectively. The displacement of each O ion is indicated by the arrow attached to the ion. The subtle differences between edge length of Re$_4$ tetrahedra responding to O ion movement are distinguished by the colors. The shaded area shows the dihedral mirror plane falling along the “indigo” and “purple” edge of Re$_4$ tetrahedra and the curved arrow shows the twofold axis through the connecting Re ion. (d) Saddle points on the energy surface from phase II to III as a function of root sum squared (rss) distance of atoms, where the energy of phase II has set to zero.}
     \label{fig:distortion}
\end{figure}

The non-centrosymmetric lattice, strong spin-orbit coupling and metallic properties of the two well-characterized low-temperature phases of Cd$_2$Re$_2$O$_7$ naturally satisfy the prerequisites for non-magnetic Weyl semimetals, providing an ideal platform to study the movement of Weyl nodes constrained by crystallographic symmetry changes. This inspires us to consider a smooth path connecting phases II and III. The crystal structures of phases II and III are obtained through a first-principles phonon calculation performed relativistically (including spin-orbital coupling) with a revised Perdew-Burke-Ernzerhof GGA functional for solids (PBEsol)~\cite{Perdew2008} on phase I (see details in SM), and the intermediate structures are obtained using the nudged elastic band (NEB) algorithm~\cite{Henkelman2000} to optimize a number of linear interpolated structures between phases II and III. The results show that all fully optimized intermediate structures belong to the same space group, constituting a smooth transition path of F222 symmetry. Including the start and end of the path (namely phases II and III), any crystal structure on the path can be interpreted as a consequence of structural distortions of phase I, which is dominated by Cd and O ion displacements. Figs. \ref{fig:distortion}(b-c) visualize the distortions of O ions occurring in phases II and III. It is worth noting that two neighbouring O$_6$ octahedra have displacements with opposite directions, thus the inversion symmetry of the system is broken. Interestingly, although the distortion does not have a Jahn-Teller origin but instead is very likely driven by strong spin-orbital coupling~\cite{Fu2015}, it exhibits extremely similar behavior compared to the Jahn-Teller effect occurring among $d$-orbital coordination complexes in which $E_g$ electronic states are coupled to $E_g$ vibrational modes. In phase II, the distortion conserves the symmetry of the horizontal square of the O$_6$ octahedra and gives half of the O$_6$ octahedra an elongation along the vertical direction whereas the other half a compression. In phase III, the horizontal square is distorted to a rhombus, which breaks the vertical mirror planes that fall along the “indigo” and “purple” edges of Re$_4$ tetrahedra shown in Fig. \ref{fig:distortion}(b). It is also worth noting that in phase II the two neighbouring Re$_4$ tetrahedra are non-congruent, however when it comes to phase III, they become congruent via losing the mirror plane in phase II and adding a twofold axis through the connecting Re ion, as shown in Fig. \ref{fig:distortion}(c)). Such a change in symmetry results in the conversion of the point group of the system from D$_{2d}$ to D$_4$. We show later that this transition from a fourfold rotoinversion symmetry ($S_4 = I C_{4z}^{-1}$ in D$_{2d}$) to a proper fourfold rotation symmetry ($C_{4z}$ in D$_4$) drives a major topological transformation of the band structure in reciprocal space.

As a final remark, we emphasize that the path proposed in this work is not just a fictitious computational construct. Its existence is strongly supported by group theory analysis~\cite{Sergienko2003} and by the latest magnetic torque experiment performed with extremely small temperature steps~\cite{Uji2021}. The phenomenological order parameter characterizing the cubic-to-tetragonal phase transitions of Cd$_2$Re$_2$O$_7$ corresponds to a twofold degenerate $E_u$ representation of the cubic point group, so it makes sense that the NEB calculation has found a series of orthorhombic intermediate phases of space group F222 (the maximal common subgroup of phases II and III) which are induced by a mixture of the two components of the representation. The result from the NEB calculation shows that the energy difference of intermediate phases on the path is anomalously small ($\Delta E \leq \SI{0.3}{\meV \per atom}$, see Fig \ref{fig:distortion}(d)), that is of the order of $\Delta T=$\SI{0.4}{\kelvin} for a structure traversing the whole path. This extremely flat path leads to a quasi-continuous $U(1)$ symmetry of the energy landscape of Cd$_2$Re$_2$O$_7$. Breaking this symmetry was predicted to yield a Goldstone mode manifesting as a structural fluctuation between phase II and III with vanishing frequency and excitation energy, and this has been observed by polarized Raman scattering and X-ray diffraction experiments~\cite{Kendziora2005, Venderley2020}. In particular, we have found an energy local minimum on the path which has the same energy as phase II, and we propose it as a candidate structure for the newly discovered intermediate phase in Ref.\,\cite{Uji2021}.

\begin{figure}
     \centering
     \includegraphics[width=0.5\textwidth]{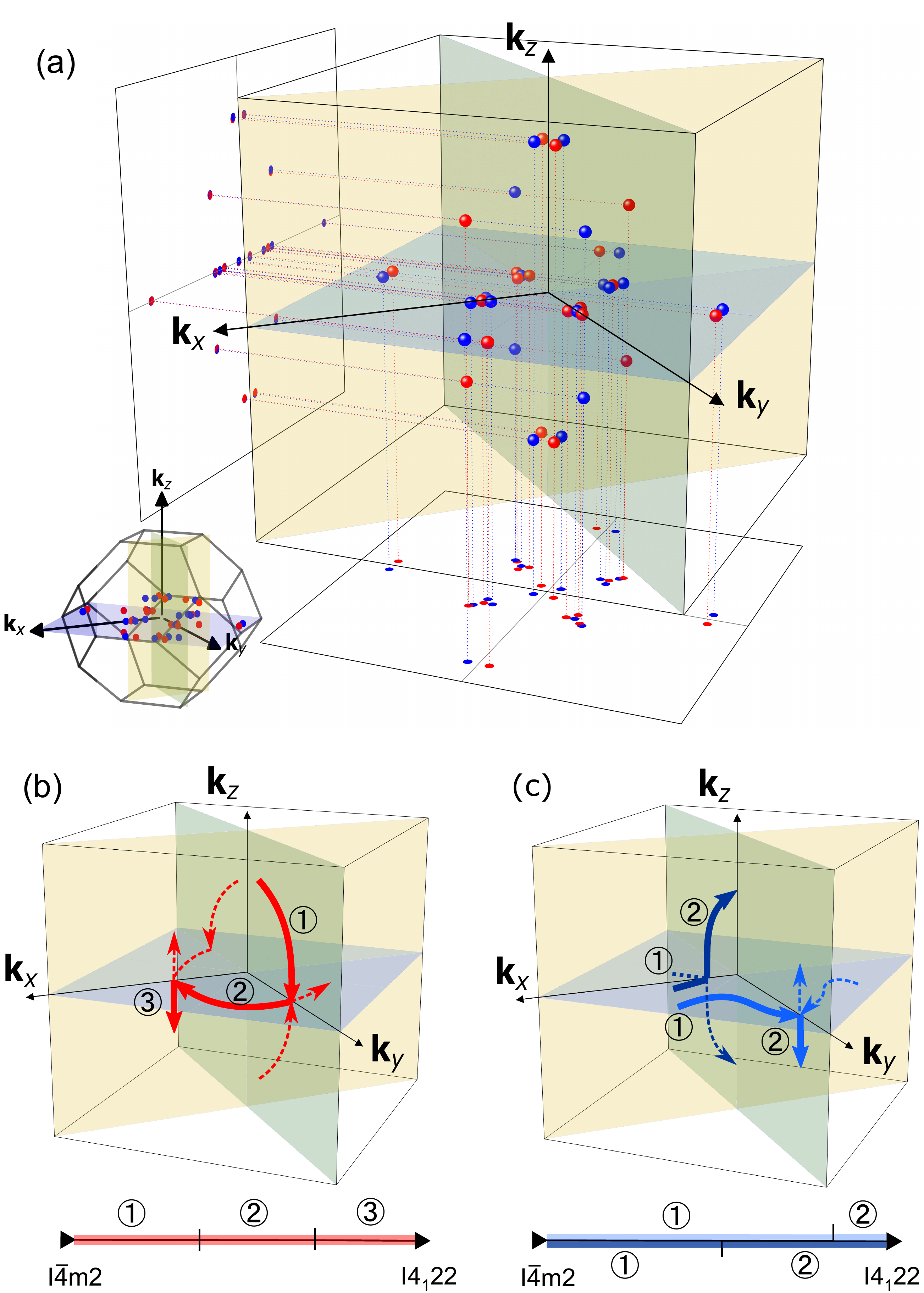}
     \caption{(a) Weyl nodes in the Brillouin zone of phase II. The chirality of the Weyl nodes is distinguished by the red and blue colors of the balls/dots. The three $C_2T$-invariant planes are highlighted. The Brillouin zone polyhedron is shown in the lower left corner for reference. (b-c) The movement of the red and the blue Weyl nodes as the phase transition smoothly progresses from phase II to phase III. The thick arrows indicate the movement of the Weyl nodes of interest, and the dash arrows indicate the movement of their symmetry-related partners. The bars below show the ``timelines" of the movement. See text for more details. }
     \label{fig:weyl}
\end{figure}

\parag{Mechanism of Weyl node movement} Since the path connecting phases II and III preserves the symmetry of space group F222, the key symmetry elements constraining the phase transition are three twofold rotation axes along the $k_x$, $k_y$ and $k_z$ directions respectively: $C_{2x}$, $C_{2y}$, and $C_{2z}$. Time-reversal symmetry $T$ is also preserved along the transition path as no magnetic ordering has been observed in experiments~\cite{Vyaselev2002} or in our first-principles calculations. Hence, there are three mutually perpendicular $C_{2}T$-invariant planes dividing the Brillouin zone polyhedron into eight equivalent regions, on which state $\left|\psi_{k}\right\rangle$ can be mapped to itself by a combination of symmetry operators $C_{2i}T$ ($i=x$ or $y$ or $z$). Fig.\ref{fig:weyl}(a) shows the distribution of the Weyl nodes of phase II formed by the two partially occupied bands closest to the Fermi energy $E_\mathrm{F}$, whose main contributions are Re $t_{2g}$ orbitals (see the band structure in SM). There are a total of twenty pairs of Weyl nodes in the whole Brillouin zone located within the energy window of $E_\mathrm{F} \pm \SI{100}{\meV}$. Considering the symmetry of the system, there are four independent octets and two independent quartets of Weyl points. All of them are pinned exactly on the $C_2T$-invariant planes, which is in line with the previous theoretical prediction that $C_2T$ symmetry is able to stabilize Weyl nodes~\cite{Wi2,bouhon2019wilson,BzduConversion}. Additionally, due to the presence of two dihedral mirror planes ($k_x$=$k_y$ and $k_x$=$-k_y$ planes) the Weyl nodes flip chirality when mapped from one vertical $C_{2}T$-invariant plane to another, thus forming quartets of Weyl points at fixed $k_z$ with alternating chirality. $C_{2z}T$ symmetry then maps each quartet of Weyl points at $k_z$ to a quartet at $-k_z$ with no change of chirality, thus giving rise to each independent octet of nodes in Fig.~\ref{fig:weyl}(a) where the $S_4$ symmetry maps each Weyl node at a fixed $k_z$ to a Weyl node of opposite chirality at $-k_z$. In contrast, the symmetries of phase III require that each quartet of nodes at a fixed $k_z$ has a unique chirality as imposed by $C_{4z}$ symmetry, and similarly for the quartet mapped to $-k_z$ by $C_{2z}T$, thus giving rise to octets of nodes of the same chirality. While the transition from one phase to another requires the splitting of the octets into D$_2$ symmetric quartets of the same chirality (each contained in one of the two vertical $C_2T$ planes), the intrinsic incompatibility between the D$_{2d}$-symmetric octets of alternating chirality and the D$_4$-symmetric octets of the same chirality enforces a qualitative rearrangement of the Weyl nodes through the Brillouin zone across the structural phase transition. It is worth noting that unlike the situation in real space where each atom can move along three independent degrees of freedom, Weyl nodes pinned on the $C_2T$-invariant planes in reciprocal space only have two degrees of freedom, as a single Weyl node on a given $C_2T$-invariant plane cannot get out of that plane unless it recombines with another Weyl node. Therefore, the Weyl nodes are forced to move a large distance across the Brillouin zone as the structural phase transition occurs, despite the very small temperature change and associated small atomic displacements in real space.

The trajectory of the Weyl nodes with `red' chirality on the $C_{2x}T$-invariant plane (shown in Fig.\ref{fig:weyl}(b)) is a typical example reflecting the above scenario. In the initial configuration the Weyl nodes on the two vertical $C_{2i}T$-invariant planes ($i=x,y$) are mirror images of each other. When the mirror symmetry breaks and the system gradually approaches phase III with fourfold rotational symmetry, the `red' chirality Weyl nodes on the $C_{2x}T$-invariant plane with $k_y>0$ first move towards each other and meet at the $C_{2z}T$-invariant plane. After the collision, they get out of the $C_{2x}T$-invariant plane and continue moving on the $C_{2z}T$-invariant plane in opposite directions, until each of them meets its previous $C_{2z}$-rotational symmetry partner with $k_y<0$ at the intersection of the $C_{2y}T$- and $C_{2z}T$-invariant planes. Then they leave the horizontal $C_{2z}T$-invariant plane and stop somewhere on the $C_{2y}T$-invariant plane. To summarize, in order to migrate from one vertical $C_2T$-invariant plane to another, the `red' Weyl nodes have to travel nearly a quarter of the Brillouin zone. Fig.\ref{fig:weyl}(c) shows another two examples of significant motion of Weyl nodes. The two `blue' chirality Weyl nodes located near the boundary of the $C_{2z}T$-invariant plane meet at the $k_y$ axis and finally stop somewhere on the $C_{2x}T$-invariant plane (indicated by the light blue arrows). The two `blue' chirality Weyl nodes that are located on the $C_{2z}T$-invariant plane and initially very close to each other first slowly meet at the $k_x$ axis, then shift over an arc on the $C_{2y}T$-invariant plane approaching the $k_z$ axis (indicated by the dark blue arrows). 

Due to the bulk-boundary correspondence, the movement of bulk Weyl nodes can also be traced in terms of the surface Fermi arcs. Figs.\ref{fig:arc} (a-b) show the surface states of phases II and III on the [001] surface. We note that phase II exhibits an ideal Weyl semimetal feature --- two thick and straight Fermi arcs connecting the projection of $C_{2x}T$- and $C_{2y}T$-invariant planes, which should be easily detected by angle-resolved photoemission spectroscopy. To reach the fourfold rotational symmetry in phase III, the Fermi arcs in phase II must be broken, therefore the Fermi arcs in phase III are more local and not as obvious as in phase II.
 
\begin{figure}
     \centering
     \includegraphics[width=0.5\textwidth]{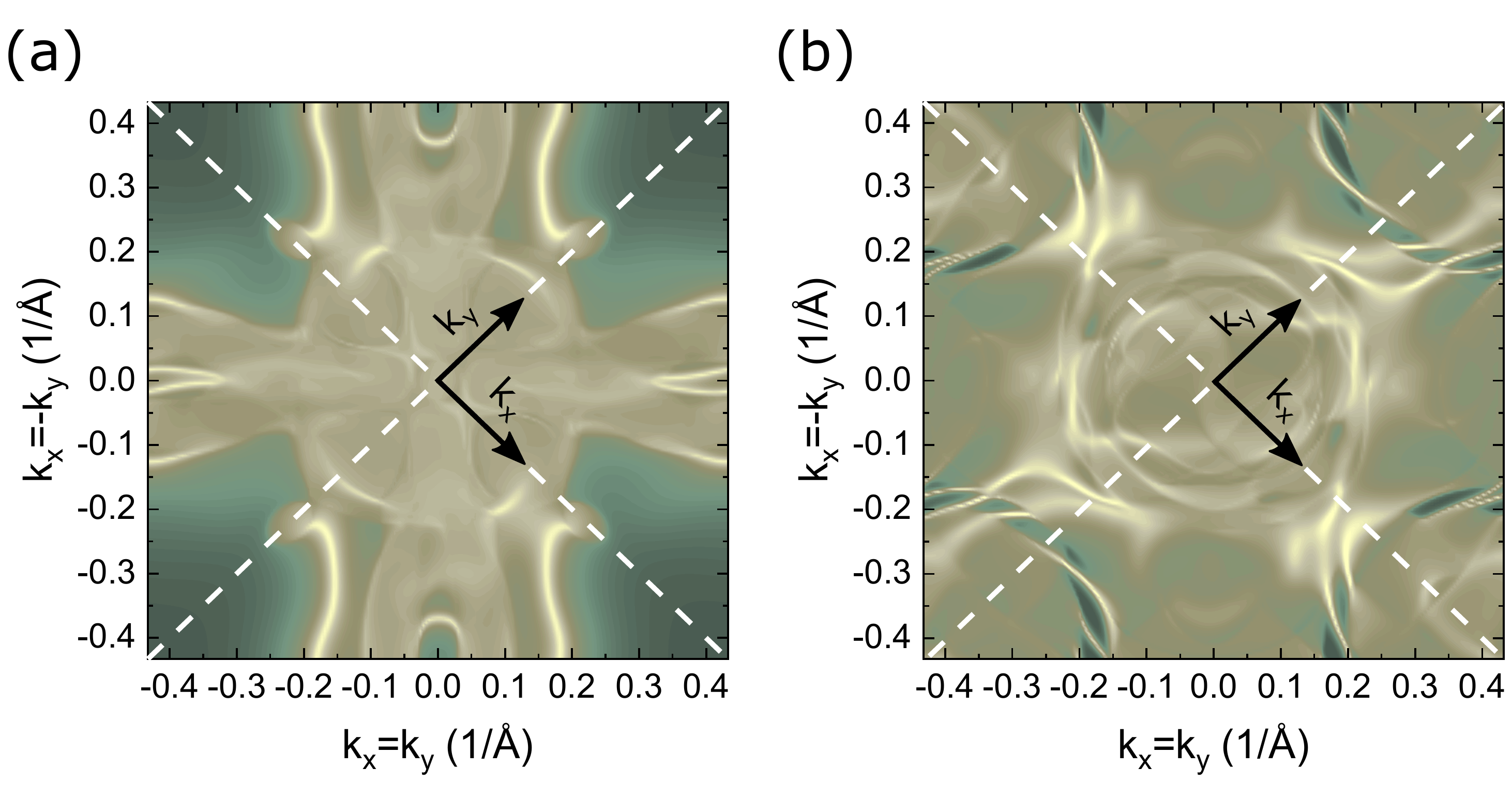}
     \caption{$\boldsymbol{k}$-dependent local density of states (LDOS) for phases (a) II and (b) III on the [001] surface, where gold (olive) colour represents high (low) LDOS. The diagonal dash lines highlight the projections of $C_{2x}T$- and $C_{2y}T$-invariant planes on the [001] surface.}
     \label{fig:arc}
\end{figure}

\parag{$C_2$-indicated locking of Weyl nodes and unlocking through braiding} Crucially, we find that a subset of pairs of Weyl nodes are locked on their $C_2T$ plane as a consequence of non-trivial Euler class topology~\cite{Ahn2019, bouhon2019nonabelian, bouhonGeometric2020}. This implies that the braiding with {\it adjacent} nodes, that is, nodes within the next energy gap below or above, must take place before their transfer to another $C_2T$ plane. Calling the Weyl nodes considered so far the {\it principal} nodes, we take as an example the $C_{2x}$-symmetric pair of Weyl nodes with the same (`blue') chirality located within the horizontal $C_{2z}T$ plane at $k_x>0$ (whose trajectories are indicated by the dark blue arrows in Fig.~\ref{fig:weyl}(c)). We have computed the Euler class $\chi_{C_{2z}T}$ on a patch containing this pair of principal Weyl nodes in phase II, and find $\chi_{C_{2z}T}=1$ (see details in SM). The non-trivial Euler class indicates an equal $C_{2z}T$-non-Abelian frame charge of the two nodes~\cite{Ahn2019, bouhon2019nonabelian, Wu1273, bouhonGeometric2020,BzduConversion}, which implies a topological obstruction to transferring these nodes to the vertical $C_{2y}T$ plane. During the structural phase transition, we observe a band inversion happening in the adjacent gap below, which creates a new pair of adjacent nodes on the $\Gamma X$-line that crosses the patch of the principal nodes (see the snapshots in SM), that is, the principal nodes are braided with the adjacent ones. This leads to the trivialization of the $C_{2z}T$-Euler class for the principal nodes on the patch, which then allows them to move to the vertical $C_{2y}T$ plane after their merging on the $C_{2x}$-symmetric $\Gamma X$-line. It is remarkable that the non-vanishing of the $C_{2z}T$-Euler class in phase II readily follows from the opposite nature of the $C_{2x}$-eigenvalues of the bands forming the principal nodes, see also Ref.~\cite{Murakamie1602680}. In this regard, the unlocking of the nodes from the $C_{2z}T$ plane is indicated by the exchange of the $C_{2x}$-eigenvalue of the lower band, which is induced by the braiding through an adjacent band inversion, so that both bands now have equal $C_{2x}$-eigenvalue. We conclude that on top of inducing large trajectories of the Weyl nodes across the Brillouin zone, the structural phase transition in Cd$_2$Re$_2$O$_7$ also requires multiple interesting braiding processes of Weyl nodes that are necessary to unlock their transfer from one $C_2T$ plane to another. 

\parag{Generalizations} We emphasize that the physical manifestations addressed in this work are controlled by symmetry, and not by the details of the specific material involved. Other materials have also been shown to undergo a structural phase transition mediated by a Goldstone mode, such as the manganites~\cite{Skjaervo2019, Meier2020, Juraschek2020, Marthinsen2018}. A common feature of these systems is the breaking of the $U(1)$ symmetry of their energy landscape (as an effect the discrete point group symmetry of the crystal) with discrete minima separated by small potential barriers. While in our case this facilitates an unconventional phase transition between two isomorphic but distinct crystal structures, in the case of the manganites, the high-temperature phase $P6_3/mmc$ ($D_{6h}$) gives rise to two non-isomorphic lower-temperature phases $P6_3cm$ ($C_{6v}$) and $P\bar{3}c1$ ($D_{3d}$), that both exhibit a sixfold degenerate ground state. Focusing on the non-centrosymmetric phase ($P6_3cm$) which is the only one permitting Weyl nodes from a symmetry perspective, each discrete minimum corresponds to a distinct polar lattice distortion (combining three Mn-trimerizations and two ferroelectric polarizations)~\cite{Skjaervo2019, Meier2020, PhysRevLett.113.267602, PhysRevB.85.174422}. This suggests that our framework would also apply in the case of a transition between two distinct polar phases. These questions, as well as studies into other symmetry groups and materials, provide intriguing future research directions. 

\parag{Conclusions and discussion}
In conclusion, we have shown that phase transitions in the materials science sense can result in relatively large movements of Weyl nodes due to symmetry changes across the transition. This offers an interesting route to significantly manipulate Weyl nodes and their associated Fermi arcs using only moderate temperature changes. Although the principles behind the mechanism are generic, we have also discussed a specific material example in the form of Cd$_2$Re$_2$O$_7$. We note a recent high-throughput screening for Weyl semimetals with fourfold rotoinversion symmetry~\cite{GAO2021}, and based on our findings it is worthwhile to explore whether these candidates may have any phase transition from $S_4$ to $C_4$, which would suggest similar physics to that reported here. Overall, our work sets the stage for the practical manipulation Weyl nodes, a prerequisite for their ultimate technological use. \\ 

\parag{Acknowledgements}
S.C. acknowledges financial support from the Cambridge Trust and from the Winton Programme for the Physics of Sustainability. R.-J.~S. acknowledges funding from the Marie Sk{\l}odowska-Curie Programme under EC Grant No. 842901, the Winton Programme for the Physics of Sustainability, and Trinity College at the University of Cambridge. B.M. acknowledges support from the Gianna Angelopoulos Programme for Science, Technology, and Innovation and from the Winton Programme for the Physics of Sustainability. The calculations in this work have been performed using resources provided by the Cambridge Tier-2 system (operated by the University of Cambridge Research Computing Service and funded by EPSRC [EP/P020259/1]), as well as by the UK Materials and Molecular Modelling Hub (partially funded by EPSRC [EP/P020194]), Thomas, and by the UK National Supercomputing Service, ARCHER. Access to Thomas and ARCHER was obtained via the UKCP consortium and funded by EPSRC [EP/P022561/1]).



\bibliographystyle{apsrev4-1}
\bibliography{references}

\pagebreak
\onecolumngrid
\begin{center}
\textbf{\large Supplemental material \textit{for} Manipulation and braiding of Weyl nodes using symmetry-constrained phase transitions}
\end{center}
\setcounter{equation}{0}
\setcounter{figure}{0}
\setcounter{table}{0}
\setcounter{page}{1}
\renewcommand{\theequation}{S\arabic{equation}}
\renewcommand{\thefigure}{S\arabic{figure}}

\title{Supplemental Material for\\ Manipulation and braiding of Weyl nodes using symmetry-constrained phase transitions}

\author{Siyu Chen}
\affiliation{TCM Group, Cavendish Laboratory, University of Cambridge,
J. J. Thomson Avenue, Cambridge CB3 0HE, United Kingdom}
\author{Adrien Bouhon}
\affiliation{Nordic Institute for Theoretical Physics (NORDITA),
Stockholm, Sweden}
\author{Robert-Jan Slager}
\affiliation{TCM Group, Cavendish Laboratory, University of Cambridge,
J. J. Thomson Avenue, Cambridge CB3 0HE, United Kingdom}
\author{Bartomeu Monserrat}
\affiliation{TCM Group, Cavendish Laboratory, University of Cambridge,
J. J. Thomson Avenue, Cambridge CB3 0HE, United Kingdom}
\affiliation{Department of Materials Science and Metallurgy, University of Cambridge, 27 Charles Babbage Road, Cambridge CB3 0FS, United Kingdom}

\maketitle

\section{Phonons in phase I}
\begin{figure}[!h]
  \includegraphics[width=\linewidth]{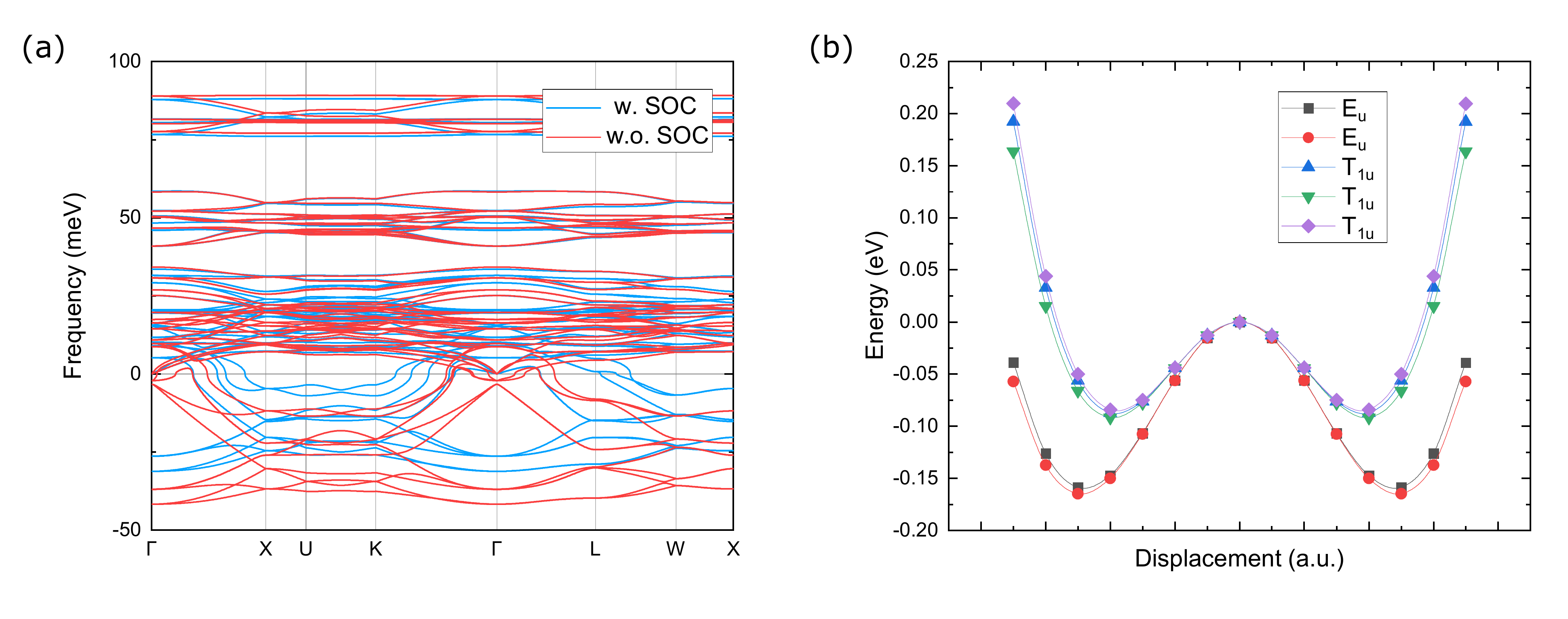}
  \caption{(a) Phonon dispersions of phase I of Cd$_2$Re$_2$O$_7$ with and without SOC. (b) The energy of phase I as a function of the frozen soft phonon amplitudes, where the energy of phase I has been set to zero.}
  \label{fig:phonon}
\end{figure}

The crystal structure of Cd$_2$Re$_2$O$_7$ at room temperature has been reported in entry 28445 of the Inorganic Crystal Structure Database~\citelatex{sm-Mariette2004} with space group Fd$\bar{3}$m. Starting from this experimentally reported structure, we have first performed a symmetry-constrained geometry optimization using the Vienna \textit{Ab initio} Simulation Package ({\sc vasp})~\citelatex{sm-VASP1,sm-VASP2}, followed by a phonon calculation with and without considering spin-orbital coupling (SOC) using the finite displacement method in conjunction with nondiagonal supercells~\citelatex{sm-Lloyd2015}. The results taking SOC into account (blue curves in Fig.~\ref{fig:phonon}(a)) show two soft modes at the center of the Brillouin zone, a twofold degenerate $E_u$ mode and a threefold degenerate $T_{1u}$ mode. Following the soft phonons, we have found that the two-fold degenerate $E_u$ phonon mode drives the structure to the I$\bar{4}$m2 and I4$_1$22 phases (which are phases II and III observed in experiments), and the three-fold degenerate $T_{1u}$ mode drives the structure to R3m, Ima2, and Cm phases, which have relatively higher energy compared to phases II and III (see Fig.~\ref{fig:phonon}(b)). Also, by comparing the phonon dispersions calculated with and without SOC, we have noticed that SOC suppresses the distortions caused by the soft modes to a certain extent. The mechanism behind this is the same as what has been found in In$_5$Bi$_3$, where the SOC interaction reduces the density of states near the Fermi energy and stabilizes the structure~\citelatex{sm-Chen2019}.

\section{Band structure of phase II}
\begin{figure}[!h]
  \includegraphics[width=0.7\linewidth]{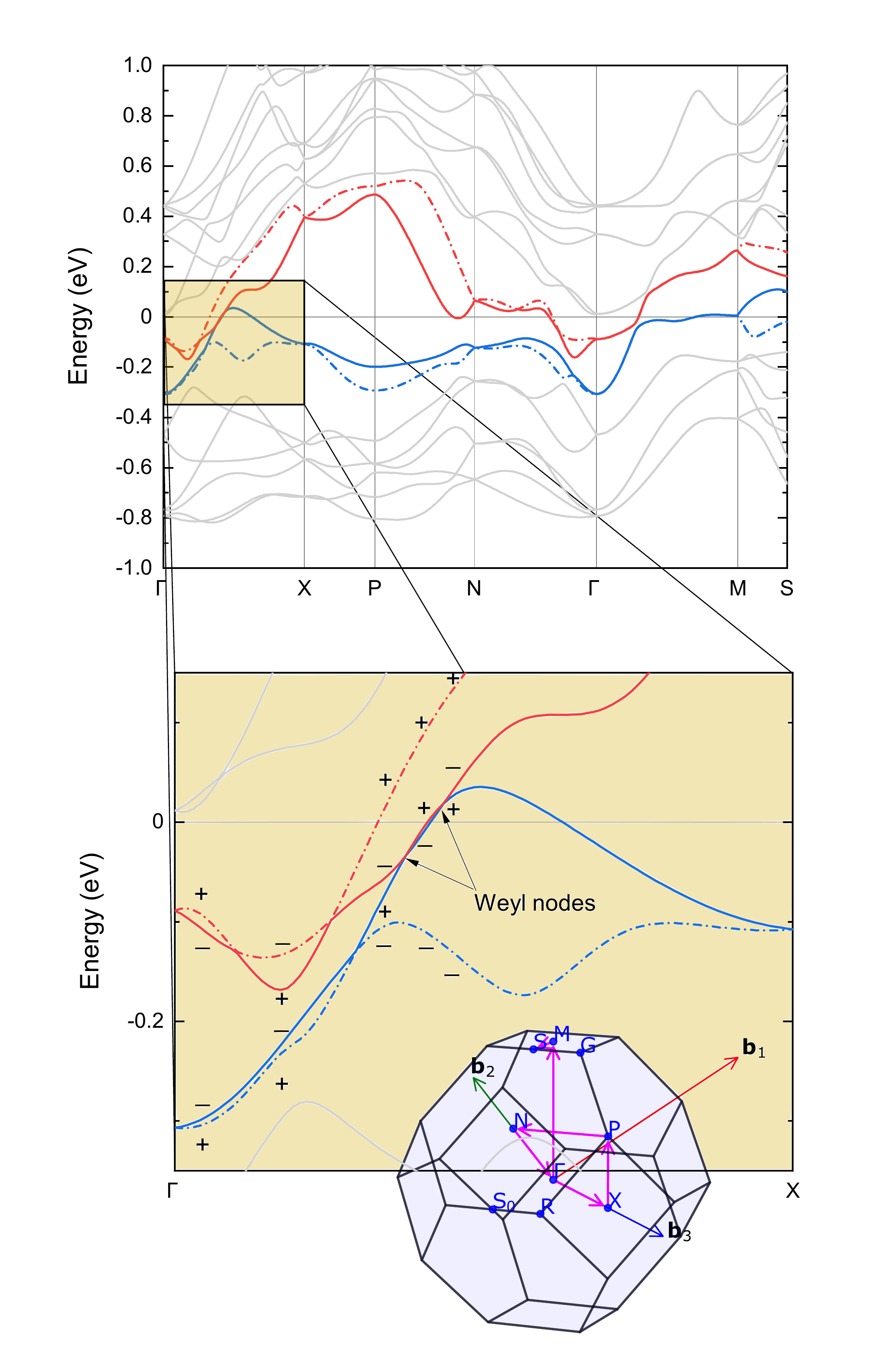}
  \caption{Band structure of phase II of Cd$_2$Re$_2$O$_7$ with SOC. The shaded region is zoomed in to show the details of the band crossings, where the `$+$' sign and `$-$' sign mark the $C_{2x}$ eigenvalues of the bands. The corresponding high-symmetry path is shown in the Brillouin zone polyhedron.}
  \label{fig:band}
\end{figure}

The band structure of phase II is shown in Fig.~\ref{fig:band}, where four partially occupied bands closest to the Fermi energy (which has been set to zero) are highlighted. The red and blue solid (dashed) bands are the $\mathcal{B}_{\pm1}$ ($\mathcal{B}_{\pm 2}$) bands defined in the main text. Since bands $\mathcal{B}_{\pm1}$ have opposite $C_{2x}$ eigenvalues on the $\Gamma X$-line, their crossings (marked in Fig.~\ref{fig:band}), which correspond the pair of the Weyl node of opposite chirality located on the $k_x$ axis shown in Fig.~\ref{fig:weyl}(a)), are protected by symmetry.

\begin{figure}[!h]
  \begin{tabular}{ccc}
  \includegraphics[width=0.3\linewidth]{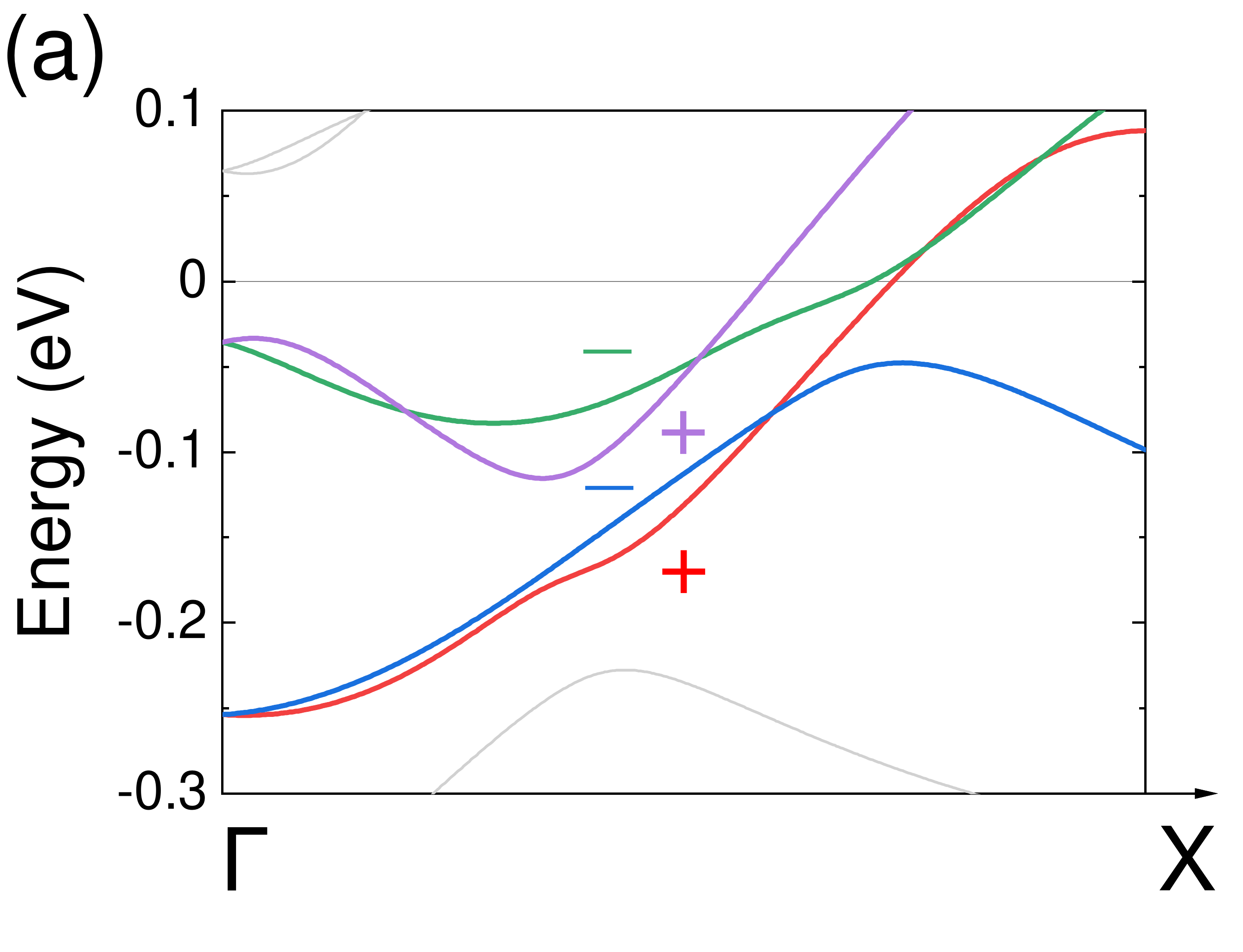} & 
  \includegraphics[width=0.3\linewidth]{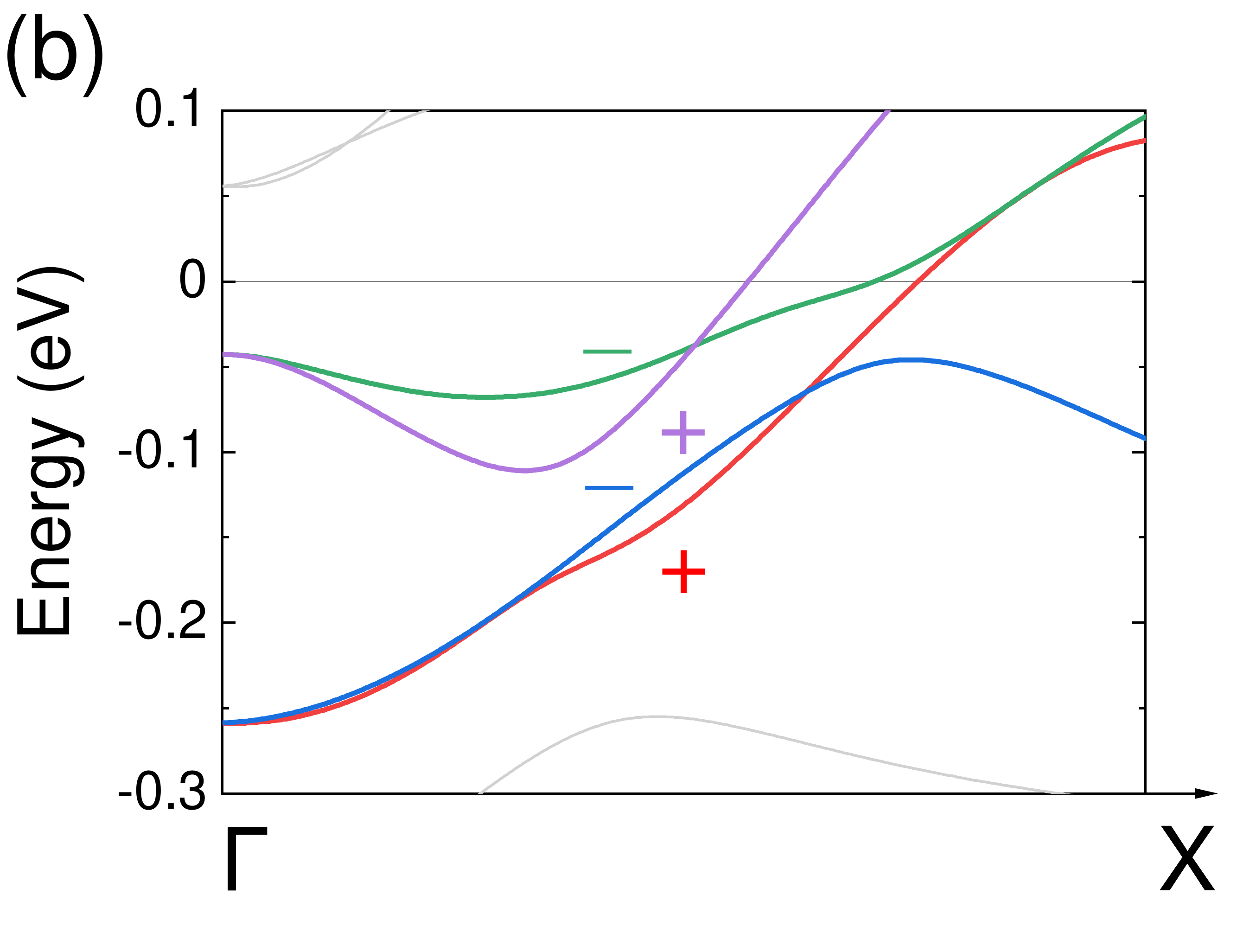} & 
  \includegraphics[width=0.3\linewidth]{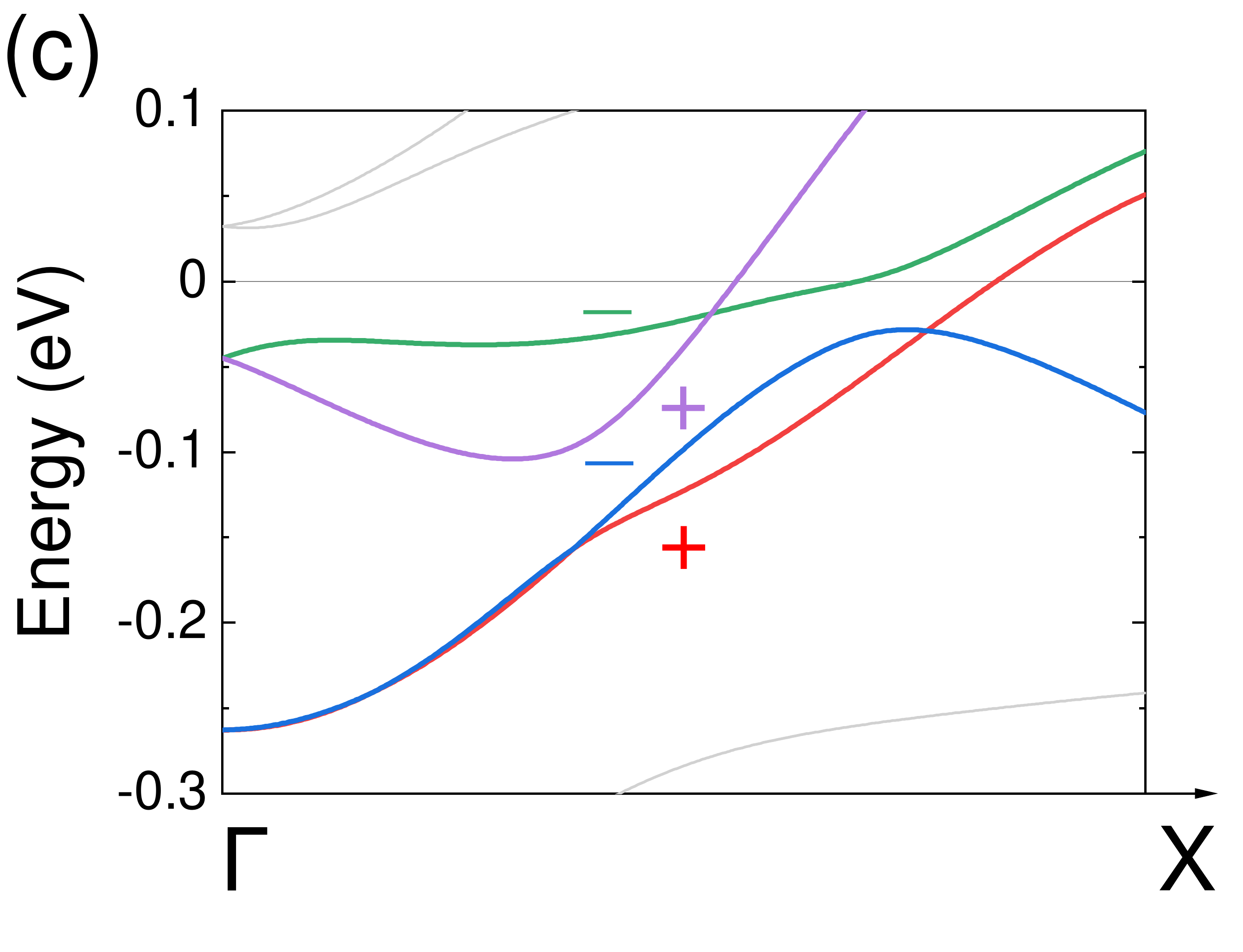} \\
  \includegraphics[width=0.3\linewidth]{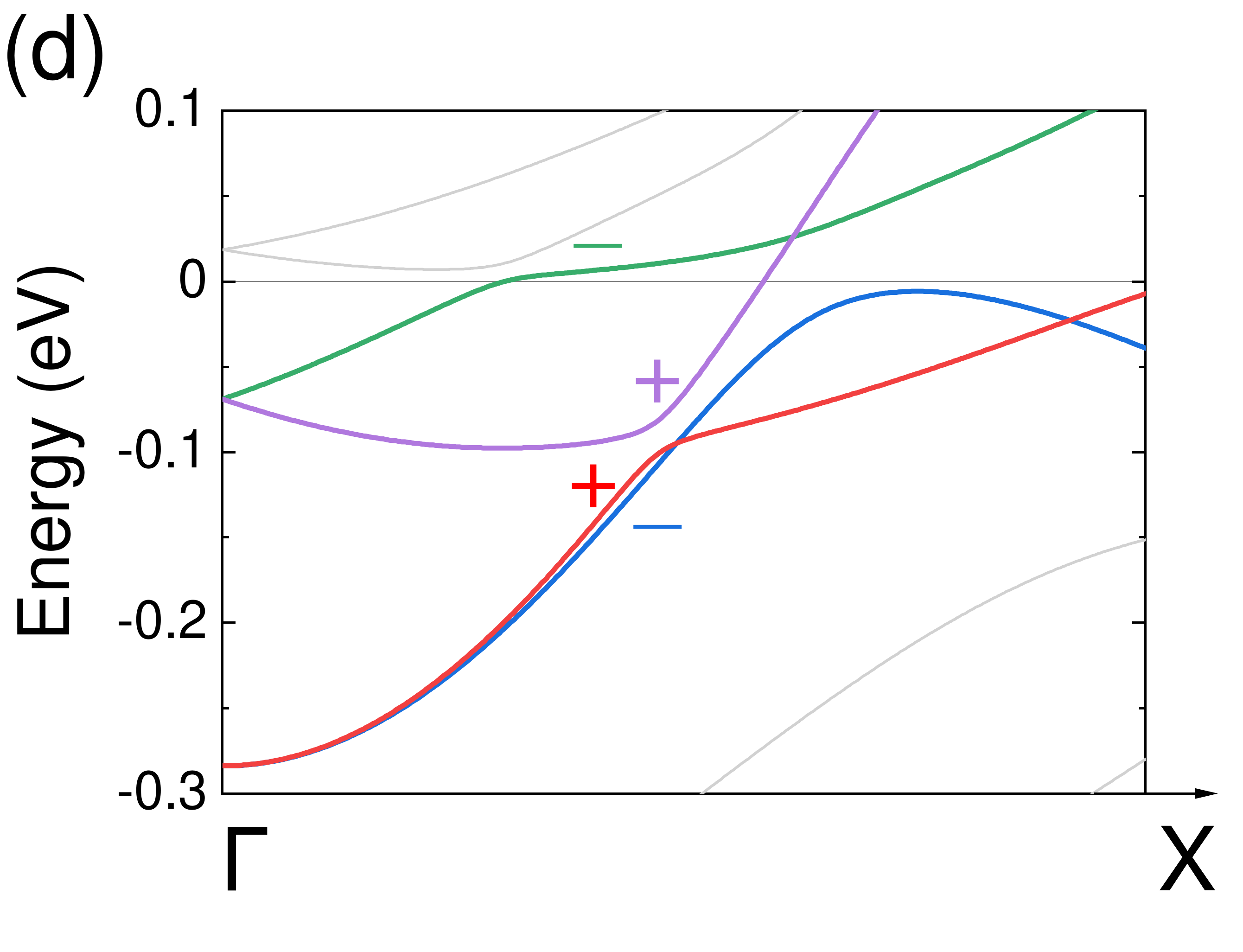} & 
  \includegraphics[width=0.3\linewidth]{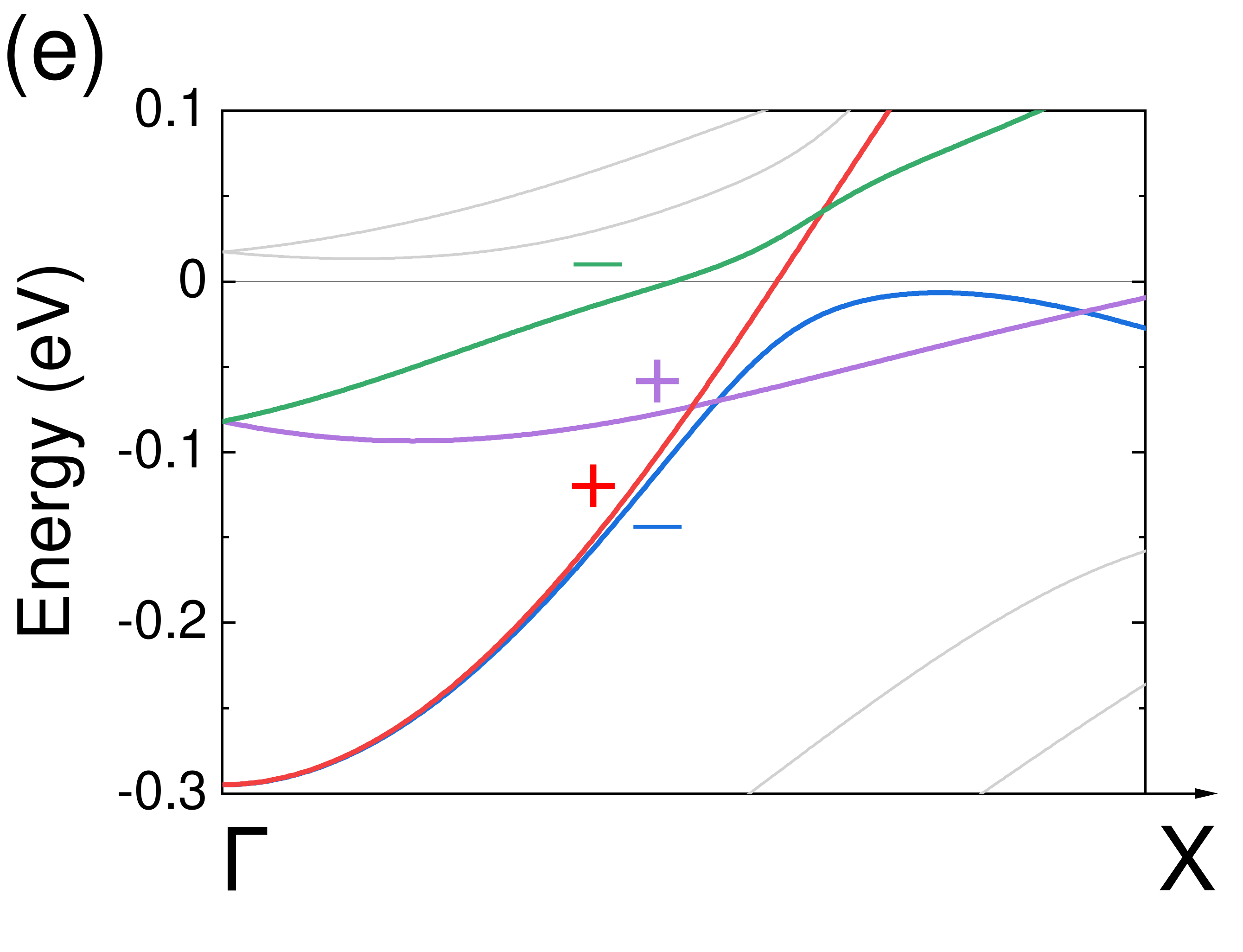} & 
  \includegraphics[width=0.3\linewidth]{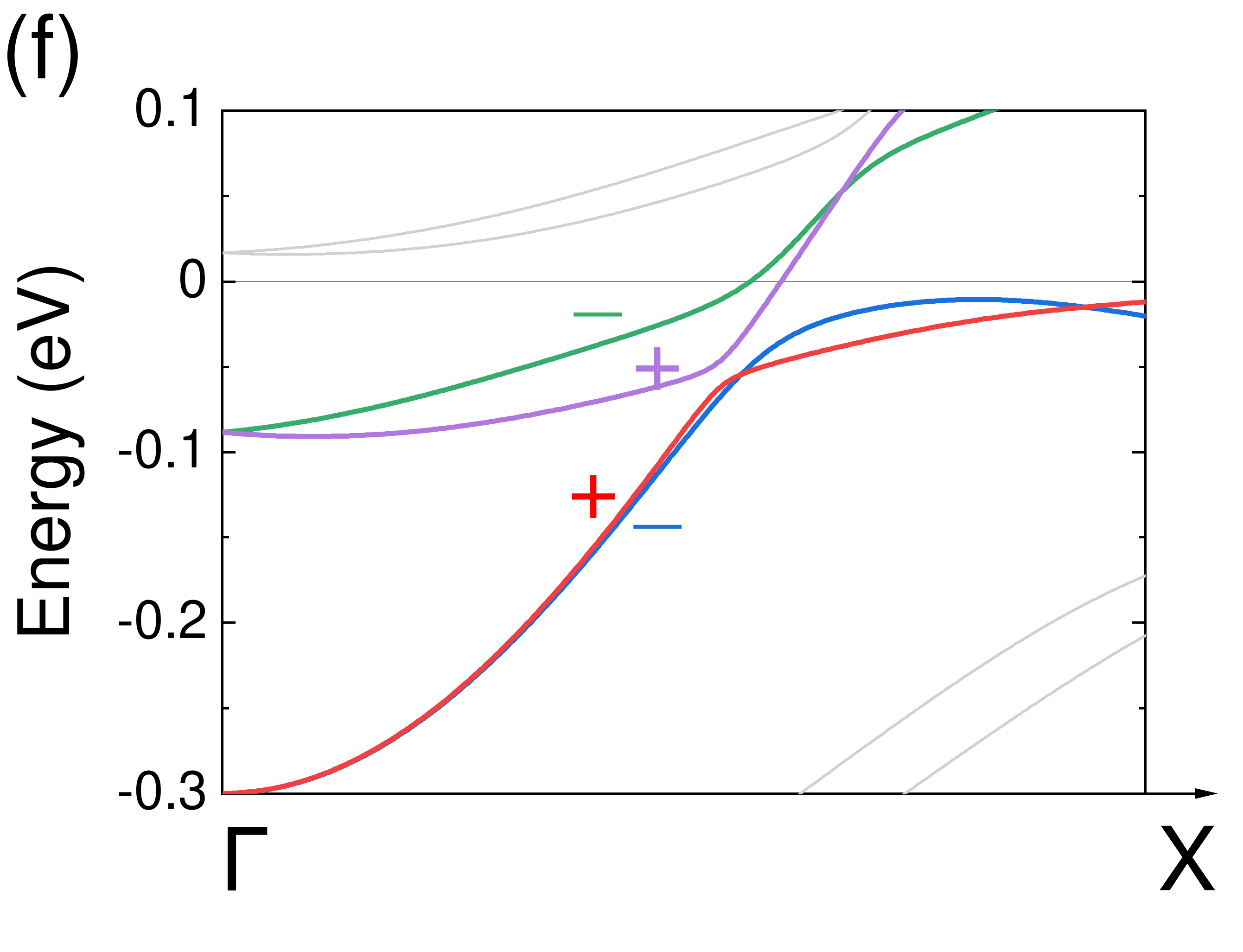}
   \end{tabular}
  \caption{(a-f) Snapshots of the band structure along the $\Gamma X$-line during the structural phase transition. The `$+$' sign and `$-$' sign mark the $C_{2x}$ eigenvalues of the bands with the same color as the symbol. }
  \label{fig:braidingsm}
\end{figure}

\section{Locking and unlocking of Weyl nodes through braiding}

We further elaborate on the Weyl node locking on $C_2T$-planes. As stated in the main text, we find that a subset of Weyl node pairs are locked on their $C_2T$ plane as a consequence of non-trivial Euler class topology~\citelatex{sm-Ahn2019, sm-bouhon2019nonabelian, sm-bouhonGeometric2020}. We call the Weyl nodes discussed in the main text the {\it principal} nodes ({\it i.e.}~formed by bands $\mathcal{B}_{-1}$ and $\mathcal{B}_{+1}$, where we label them taking 0 at the Fermi energy), and the nodes between bands $\mathcal{B}_{-1}$ and $\mathcal{B}_{-2}$ ($\mathcal{B}_{+1}$ and $\mathcal{B}_{+2}$), the lower {\it adjacent} nodes (the higher {\it adjacent} nodes). Taking as an example the same (blue) chirality and $C_{2x}$-symmetric pair of principal Weyl nodes located within the horizontal $C_{2z}T$ plane at $k_x>0$ whose trajectories are indicated by the dark blue arrows in Fig.~\ref{fig:weyl}(c) of the main text, we find that a braiding with Weyl nodes of one of the adjacent gaps is necessary to allow their transfer to the $C_{2y}T$ vertical plane. We show below that this is a consequence of the opposite $C_{2x}$-eigenvalues of the bands forming the two principal nodes. 

\begin{figure}[!h]
  \includegraphics[width=0.6\linewidth]{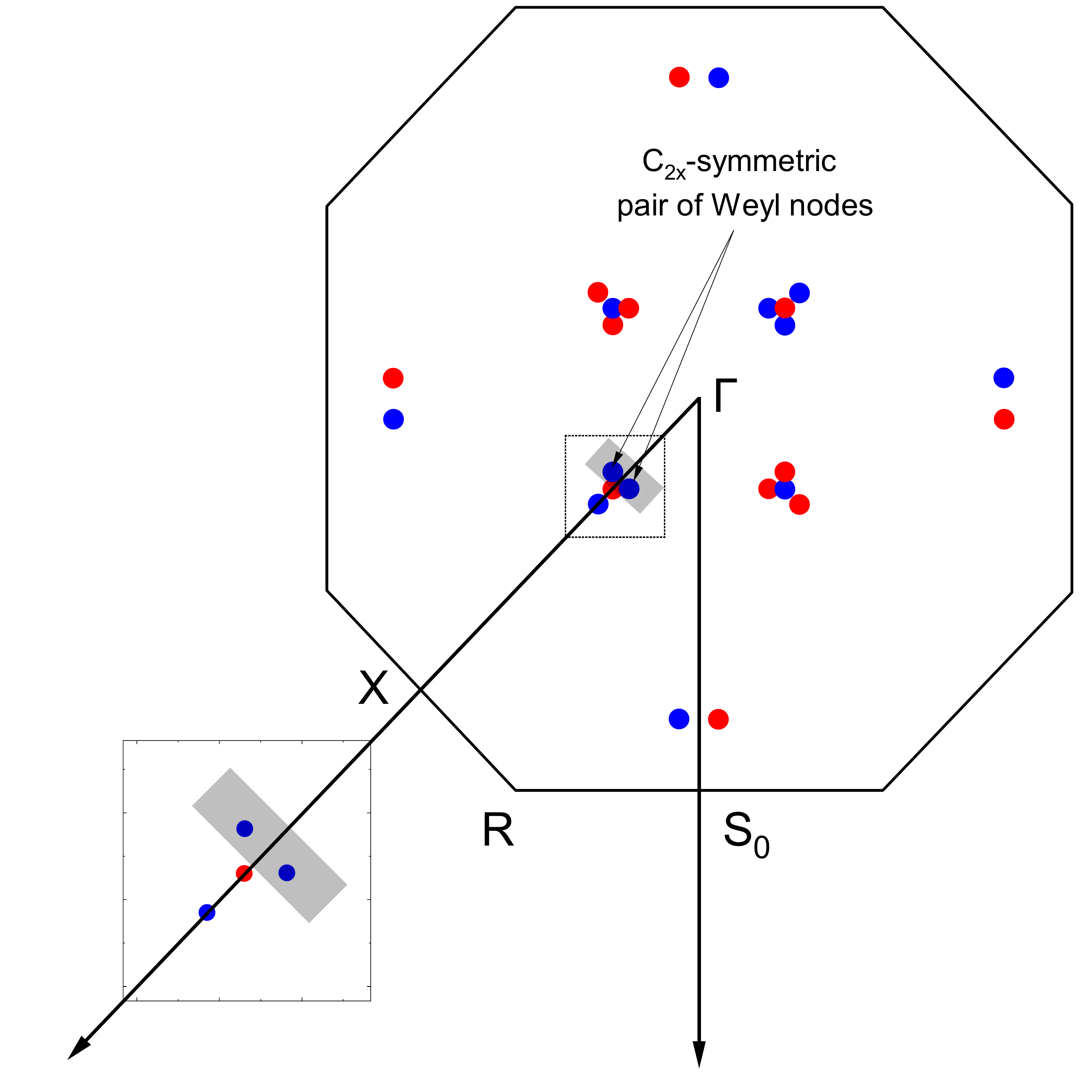} 
  \caption{Patch (marked in gray shade) of the $C_{2z}T$-invariant plane over which we compute the Euler class $\chi_{C_{2z}T}$ in phase II. The inset zooms in on the details around the $C_{2x}$-symmetric pair of Weyl nodes. Using the Bloch eigenvectors obtained through the symmetrized Wannierization, we get $\chi_{C_{2z}T}=1$, in agreement with the $C_{2x}$-eigenvalues indicator. See text for more details.}
  \label{fig:patch}
\end{figure}

Let us first consider what would happen without braiding. In phase II, the bands forming the principal nodes have opposite $C_{2x}$-eigenvalues with $\xi_{C_{2x}}[\mathcal{B}_{-1}] = -1$ and $\xi_{C_{2x}}[\mathcal{B}_{+1}] = 1$, see Fig.~\ref{fig:braidingsm}(a). In the absence of braiding, the merging of the principal nodes on the $\Gamma X$-line would induce a band inversion along the $C_{2x}$-axis between two bands of opposite $C_{2x}$-eigenvalues, forming two unavoided band crossings on the $\Gamma X$-line that are protected by $C_{2x}$, see also Ref.\,\citelatex{sm-Murakamie1602680}. As a consequence, the two principal nodes would scatter on the $\Gamma X$-line, thus remaining on the $C_{2z}T$ plane after their merging. Such a scattering of node pairs confined on a $C_{2i}T$ plane indicates a non-trivial $C_{2i}T$-Euler class topology~\citelatex{sm-bouhon2019nonabelian}. We have computed the Euler class on a patch of the $C_{2z}T$ plane that contains the two principal nodes under consideration while avoiding all other nodes, see Fig.~\ref{fig:patch}, for which we obtain $\chi_{C_{2z}T}=1$ (see below for details). This confirms the prediction based on the $C_{2x}$-eigenvalues. 

We thus conclude that a braiding with adjacent nodes must take place~\citelatex{sm-bouhon2019nonabelian} in order to unlock the pair of principal nodes from the $C_{2z}T$-plane, thus allowing its transfer on the vertical $C_{2y}T$ plane. 
Indeed, as shown in Fig.~\ref{fig:braidingsm}(a-f) with successive snapshots of the band structure during the phase transition, we find that a band inversion within the lower adjacent gap is taking place along the $\Gamma X$-line, leading to the exchange of the $C_{2x}$-eigenvalue of band $\mathcal{B}_{-1}$ and giving rise to two lower adjacent nodes along the $C_{2x}$-line. We then observe that the two principal nodes merge on the $\Gamma X$-line, which now involves two bands of equal $C_{2x}$-eigenvalue, {\it i.e.}~$\xi_{C_{2x}}[\mathcal{B}_{-1}] =\xi_{C_{2x}}[\mathcal{B}_{+1}]= 1$. As a result, the band inversion in the principal gap along the $\Gamma X$-axis now induces avoided band crossings, indicating that the two principal nodes have moved from the horizontal $C_{2z}T$ plane to the vertical $C_{2y}T$ plane. This implies that their $C_{2z}T$- and $C_{2x}T$-Euler classes on patches that avoid the adjacent nodes are both zero.

\section{Real Bloch eigenvectors}\label{ap:euler}

Performing the Wannierization of the bands around the Fermi energy over a 24-orbital-spin basis with {\sc wannier90}~\citelatex{sm-wannier90} and symmetrizing the tight-binding Hamiltonian~\citelatex{sm-PhysRevMaterials.2.103805}, we obtain the Bloch eigenvectors of the two bands forming the two principal nodes over a discrete mesh of the patch (highlighted in Fig.~\ref{fig:patch}) in momentum space. We compute the Euler class using the code provided in Ref.\,\citelatex{sm-Bzdusek:Mathematica-Euler}, for which the Bloch eigenvectors must be rotated into the basis that makes them real~\citelatex{sm-bouhon2019nonabelian}. To accomplish this, we first need to specify the effective atomic degrees of freedom of the wannierized tight-binding model, and then perform a Takagi factorization of the $C_2T$ operator acting on these.  

Let us write the primitive cell lattice vectors for space group I$\bar{4}$m2 as 
\begin{equation}
    \boldsymbol{a}_1 = \dfrac{1}{2}(-a,a,c),~
    \boldsymbol{a}_2 = \dfrac{1}{2}(a,-a,c),~
    \boldsymbol{a}_3 = \dfrac{1}{2}(a,a,-c).
\end{equation}
We obtain a Wannierized tight-binding model in the Wannier basis that spans four Re inequivalent atomic sites at the Wyckoff position $8\mathrm{i}$, with the three $d$-orbitals $\{d_{xz},d_{yz},d_{xy}\}$ and the two spin-$1/2$ components. The position vectors for four inequivalent Re sites at Wyckoff position $8\mathrm{i}$ are given through (the other four equivalent sites of Wyckoff position $8\mathrm{i}$ are obtained through a shift by $\boldsymbol{\tau} = \dfrac{1}{2}(a,a,c)$)
\begin{equation}
    \boldsymbol{r}_1 = (0,x,z),~
    \boldsymbol{r}_2 = C_{2z}\boldsymbol{r}_1=(0,-x,z),~
    \boldsymbol{r}_3 = S_{4z}\boldsymbol{r}_1=(-x,0,-z),~
    \boldsymbol{r}_4 =S^{-1}_{4z}\boldsymbol{r}_1= (x,0,-z),
\end{equation}
with the fourfold rotoinversion $S_{4z} = C_{4z}^{-1}I$. We write the Wannier-L{\"o}wdin type basis of the tight-binding model as 
\begin{equation}
    \vert \boldsymbol{\varphi}_i, \boldsymbol{R}+\boldsymbol{r}_i \rangle  =
\left(
\vert \varphi_{i,d_{xz},\uparrow} \rangle~
\vert \varphi_{i,d_{yz},\uparrow}  \rangle~
\vert \varphi_{i,d_{xy},\uparrow}  \rangle~
\vert \varphi_{i,d_{xz},\downarrow}  \rangle~
\vert \varphi_{i,d_{yz},\downarrow}  \rangle~
\vert \varphi_{i,d_{xy},\downarrow}  \rangle
\right),~\mathrm{for}~i=1,2,3,4,
\end{equation}
from which we obtain the Bloch basis per inequivalent lattice site,
\begin{equation}
    \vert \boldsymbol{\phi}_i, \boldsymbol{k} \rangle =
   \sum\limits_{\boldsymbol{R}\in \boldsymbol{T}}
    \mathrm{e}^{\mathrm{i} \boldsymbol{k}\cdot (\boldsymbol{R}+\boldsymbol{r}_i)} 
    \vert \boldsymbol{\varphi}_i , \boldsymbol{R}+\boldsymbol{r}_i \rangle
    ,~\mathrm{for}~i=1,2,3,4,
\end{equation}
with the tetragonal body-centered Bravais lattice $\boldsymbol{T} = \{n_1 \boldsymbol{a}_1,n_2 \boldsymbol{a}_2,n_3 \boldsymbol{a}_3\}_{(n_1,n_2,n_3)\in \mathbb{Z}^3}$. The total Bloch basis is then written
\begin{equation}
    \vert \boldsymbol{\Phi}, \boldsymbol{k} \rangle =
   \left( \vert \boldsymbol{\phi}_1, \boldsymbol{k} \rangle ~ \vert \boldsymbol{\phi}_2, \boldsymbol{k} \rangle ~ \vert \boldsymbol{\phi}_3, \boldsymbol{k} \rangle ~ \vert \boldsymbol{\phi}_4, \boldsymbol{k} \rangle \right).
\end{equation}
The representation of $C_{2z}T$ on the total Bloch basis is thus
\begin{equation}
    {C_{2z}T} \vert \boldsymbol{\Phi}, \boldsymbol{k}  \rangle = 
    \vert \boldsymbol{\Phi}, m_z\boldsymbol{k} \rangle \cdot U_{C_{2z}T} \mathcal{K}
\end{equation}
with the mirror operation $m_z (k_x,k_y,k_z) = (k_x,k_y,-k_z)$, the complex conjugation $\mathcal{K}$ and the $U_{C_{2z}T}$ unitary matrix given by
\begin{equation}
    U_{C_{2z}T} = \mathrm{i}\sigma_x \otimes U_{2,\mathrm{site}} \otimes U_{2,\mathrm{orb}},
\end{equation}
where $U_{2,\mathrm{site}} = \mathbb{1}_2 \otimes \sigma_x$ and $U_{2,\mathrm{orb}} = \mathrm{diag}(-1,-1,1)$.
Writing the tight-binding Bloch Hamiltonian as 
\begin{equation}
    \mathcal{H} = \sum\limits_{\boldsymbol{k}} 
    \vert \boldsymbol{\Phi}, \boldsymbol{k}  \rangle
    \cdot
    H(\boldsymbol{k})
    \cdot
    \langle \boldsymbol{\Phi}, \boldsymbol{k}  \vert 
     = \sum\limits_{\boldsymbol{k}} 
    \vert \boldsymbol{\Phi}, \boldsymbol{k}  \rangle
    \cdot
     \mathcal{V}(\boldsymbol{k})
     \cdot
    \mathcal{E}(\boldsymbol{k}) 
    \cdot
    \mathcal{V}^{\dagger}(\boldsymbol{k}) 
    \cdot\langle \boldsymbol{\Phi}, \boldsymbol{k}  \vert,
\end{equation}
with the diagonal matrix of energy eigenvalues $\mathcal{E}(\boldsymbol{k}) = \mathrm{diag}(E_1,\dots,E_{24})$ and the corresponding matrix of column eigenvectors $\mathcal{V}(\boldsymbol{k}) = (v_1 \dots v_{24})$, we obtain the real Bloch eigenvectors through a change of basis,
\begin{equation}
\begin{aligned}
    \vert \boldsymbol{\Phi} , \boldsymbol{k}  \rangle &= \vert  \widetilde{\boldsymbol{\Phi}} , \boldsymbol{k}  \rangle \cdot W, 
\end{aligned}
\end{equation}
that is
\begin{equation}
    \widetilde{\mathcal{V}}(\boldsymbol{k}) 
    = W \cdot \mathcal{V}(\boldsymbol{k}).
\end{equation}

We now show that the unitary matrix $W$ is obtained from the Takagi factorization of the $U_{C_{2z}T}$ unitary matrix. Indeed, we find $[C_{2z}T]^2 = U_{C_{2z}T} \cdot U_{C_{2z}T}^* = \mathbb{1}_{24}$, which implies $U_{C_{2z}T} = U_{C_{2z}T}^\intercal$. As for any symmetric matrix, there is a Takagi factorization given by $U_{C_{2z}T} = U_{\mathrm{tf}} \cdot \Lambda \cdot U_{\mathrm{tf}}^\intercal$, from which we define $W = \sqrt{\Lambda^*} \cdot U_{\mathrm{tf}}^{\dagger}$. The representation of $C_{2z}T$ in the new basis then gives
\begin{equation}
\begin{aligned}
    {C_{2z}T} \vert \widetilde{\boldsymbol{\Phi}}, \boldsymbol{k}  \rangle &= 
    \vert \widetilde{\boldsymbol{\Phi}}, m_z\boldsymbol{k} \rangle \cdot \left( W \cdot U_{C_{2z}T} \cdot W^\intercal\right) \mathcal{K} = \vert \widetilde{\boldsymbol{\Phi}}, m_z\boldsymbol{k} \rangle \mathcal{K}. 
\end{aligned}
\end{equation}
In the case of a unitary and symmetric matrix, as it is the case for $U_{C_{2z}T}$, the Takagi factorization can be readily obtained from the singular value decomposition $U_{C_{2z}T} = U_{\mathrm{svd}} \cdot \Lambda \cdot V_{\mathrm{svd}}$ where $\Lambda = \mathbb{1}$ for unitary matrices, through $U_{\mathrm{tf}} = U_{\mathrm{svd}} \cdot \sqrt{U_{\mathrm{svd}}^{\dagger}\cdot V_{\mathrm{svd}}^*}$~\citelatex{sm-CHEBOTAREV2014380}, from which we find $W = U_{\mathrm{tf}}^{\dagger}$.

\bibliographystylelatex{apsrev4-1}
\bibliographylatex{references}
\end{document}